\documentclass[%
 reprint,
superscriptaddress,
 amsmath,amssymb,
 aps,
]{revtex4-2}

\usepackage{graphicx}% Include figure files
\usepackage{dcolumn}% Align table columns on decimal point
\usepackage{bm}% bold math
\usepackage[version=4]{mhchem}
\usepackage{siunitx}
\usepackage{xcolor,soul}
\usepackage{float}
\usepackage{braket}
\usepackage[pdftex,hidelinks]{hyperref}

\begin{document}

\preprint{APS/123-QED}

\title{Scalable photonic integrated circuits for programmable control of atomic systems}

\author{Adrian J Menssen\textsuperscript{\textdagger}}
\author{Artur Hermans\textsuperscript{\textdagger}}%
\affiliation{Massachusetts Institute of Technology, Cambridge, MA, USA}
\author{Ian Christen}
\affiliation{Massachusetts Institute of Technology, Cambridge, MA, USA}
\author{Thomas Propson}
\affiliation{Massachusetts Institute of Technology, Cambridge, MA, USA}
\author{Chao Li}
\affiliation{Massachusetts Institute of Technology, Cambridge, MA, USA}
\author{\\Andrew J Leenheer}%
\affiliation{Sandia National Laboratories, Albuquerque, NM, USA}
\author{Matthew Zimmermann}
\affiliation{The MITRE Corporation, Bedford, MA, USA}
\author{Mark Dong}
\affiliation{Massachusetts Institute of Technology, Cambridge, MA, USA}
\affiliation{The MITRE Corporation, Bedford, MA, USA}
\author{Hugo Larocque}
\affiliation{Massachusetts Institute of Technology, Cambridge, MA, USA}
\author{Hamza Raniwala}
\affiliation{Massachusetts Institute of Technology, Cambridge, MA, USA}
\author{Gerald Gilbert}
\affiliation{The MITRE Corporation, Princeton, NJ, USA}
\author{Matt Eichenfield}
\affiliation{Sandia National Laboratories, Albuquerque, NM, USA}
\affiliation{University of Arizona, Tucson, AZ, USA}
\author{Dirk R Englund}
\affiliation{Massachusetts Institute of Technology, Cambridge, MA, USA}

\date{\today}

%Use showkeys class option 
\begin{abstract}

Advances in laser technology have driven discoveries in atomic, molecular, and optical (AMO) physics and emerging applications, from quantum computers with cold atoms or ions, to quantum networks with solid-state color centers. This progress is motivating the development of a new generation of ``programmable optical control'' systems, characterized by criteria (C1) visible (VIS) and near-infrared (IR) wavelength operation, (C2) large channel counts extensible beyond 1000s of individually addressable atoms, (C3) high intensity modulation extinction and (C4) repeatability compatible with low gate errors, and (C5) fast switching times. Here, we address these challenges by introducing an atom control architecture based on VIS-IR photonic integrated circuit (PIC) technology. Based on a complementary metal-oxide-semiconductor (CMOS) fabrication process, this Atom-control PIC (APIC) technology meets the system requirements (C1)-(C5). As a proof of concept, we demonstrate a 16-channel silicon nitride based APIC with $(5.8\pm0.4)$~\si{\nano\second} response times and $<\SI{-30}{\decibel}$ extinction ratio at a wavelength of \SI{780}{\nano\metre}. This work demonstrates the suitability of PIC technology for quantum control, opening a path towards scalable quantum information processing based on optically-programmable atomic systems. 
\end{abstract}                         
\maketitle

\begin{figure*}[htp]
     \centering
         \includegraphics[width=1\textwidth]{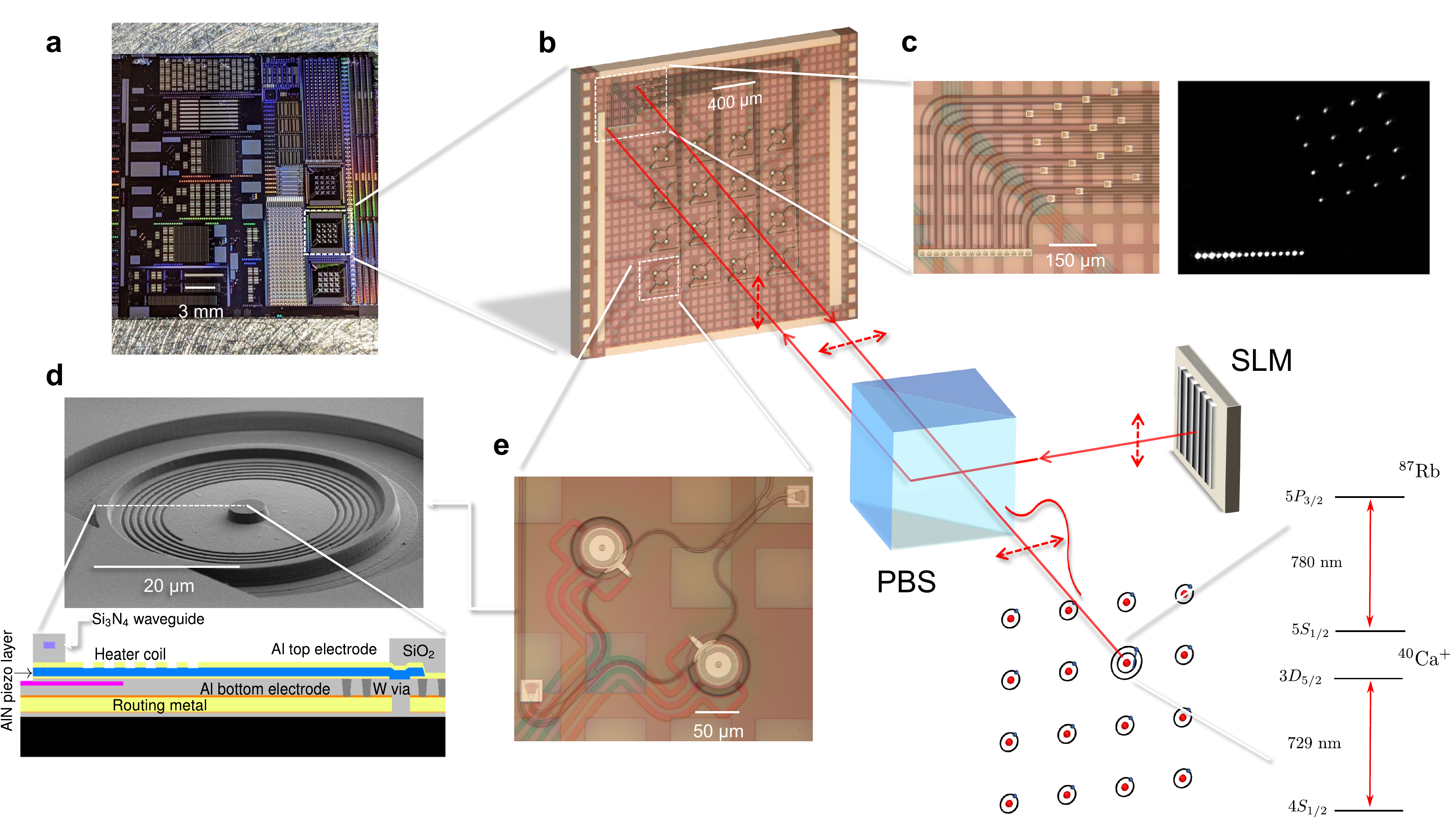}
         \caption{Atom-control photonic integrated circuit (APIC) platform. \textbf{a}, Photograph of full reticle. \textbf{b}, APIC modulator array with a modulator pitch of \SI{420}{\micro\metre}. \textbf{c}, 4$\times$4 out-coupling and 1$\times$16 in-coupling area. Chip detail (left) and camera image with light coupled into all ports (right). \textbf{d}, SEM image of individual ring (top). Schematic cross section of device with piezo-stack and waveguiding layers illustrated. \textbf{e}, DRMZM with local in- and out- coupling gratings, which can be used as an alternative to the grating couplers in \textbf{c}. Bottom right: Illustration of setup. An SLM projects light onto the APIC, where the light is modulated and after passing through a PBS imaged onto an array of (artificial) atoms. The input light path onto the SLM is not shown.}\label{fig:archi}

\end{figure*}

Quantum technologies have reached a pivotal stage where the number of qubits is approaching the limit of classical simulability~\cite{Arute:19,Zhong:20}. A key challenge in achieving practically useful quantum technology lies in the scalable, coherent control of the individual qubits.
As potential candidates for qubits, several platforms that rely on atomic or atom-like systems have emerged, including cold atoms~\cite{Graham:22,Ebadi:21}, ions~\cite{Debnath:16,Bruzewicz:19,Pino:21} and 
atom-like emitters in solids~\cite{Wan:20,Abobeih:22}. Many suitable optical transitions of these systems lie in the visible to near-infrared wavelength range. Previous work towards quantum control has relied on using a limited number of local addressing beams modulated by bulk acousto-optic devices~\cite{Omran:19,Graham:22,Debnath:16,Wright:19}. This approach becomes problematic when scaling beyond tens of optical control channels. Here, we address this bottleneck by introducing photonic integrated circuit (PIC) technology based on compact, resonant, and high extinction modulators and fabricated in a process compatible with modern complementary metal-oxide-semiconductor (CMOS) manufacturing.

The development of on-chip high-speed modulators has mainly been driven by telecom applications~\cite{Rahim:21}. However, O- and C- band platforms (e.g. silicon~\cite{Rahim:18} and indium phosphide~\cite{Smit:19}) are incompatible with visible-wavelength operation. Silicon nitride (SiN) is a leading photonic integration platform compatible with visible-wavelength operation~\cite{Rahim:17}. SiN is transparent down to blue wavelengths~\cite{Morin:21} and extremely low waveguide propagation losses have been demonstrated ($<\SI{0.1}{dB/m}$ at \SI{1.6}{\micro\metre}~\cite{Bauters:11} and \SI{22}{dB/m} at \SI{450}{\nano\metre}~\cite{Morin:21}). Moreover, SiN photonic integrated circuits are manufacturable in CMOS fabrication processes and have been demonstrated to enable high-power handling, with watt-level waveguide-coupled optical powers reported at \SI{1.6}{\micro\metre}~\cite{ElDirani:19}. While SiN's thermo-optic effect allows slow modulation with $\sim$\si{\micro\second} response times~\cite{Liang:21}, this limit in modulation rate is problematic for fast optical quantum control. Recently, SiN platforms with aluminium nitride (AlN) piezoelectric actuators have been introduced, enabling visible and near-infrared light modulation with $\sim$\si{\nano\second} response times~\cite{Stanfield:19,tian_2020_sinstiff,Dong:22}.

In this work we present an Atom-control Photonic Integrated Circuit (APIC) platform for high-fidelity local quantum control of atomic and atom-like systems, notably with high extinction (C3) and repeatability (C4).
Operating across the visible and near-infrared wavelength ranges, this platform relies on SiN photonic integrated circuits with fast AlN piezoelectric actuators~\cite{Stanfield:19}, satisfying criteria (C1) and (C5), respectively.
Our APICs are fabricated at temperatures $<\SI{500}{\celsius}$ in a \SI{200}{\milli\metre} wafer, CMOS-compatible process, permitting co-integration with electrical circuits for driving control voltages and implementing feedback~\cite{Lee:13}.
This approach enables large channel counts, thereby fulfilling criterion (C2). 

The illustration in Fig.~\ref{fig:archi} shows our proof-of-concept APIC with an array of sixteen high-speed dual-ring-assisted Mach-Zehnder modulators (DRMZMs)~\cite{Wang:07} arranged in a 4$\times$4 grid. We show that these DRMZMs can, in a fabrication-tolerant way, achieve voltage-programmable light extinction in a compact footprint, as required for high-fidelity quantum control devices with large channel counts. We demonstrate that the spread in ring resonant frequencies due to fabrication non-uniformity, a major issue in large-scale PICs~\cite{Bogaerts:14}, can be eliminated using integrated thermo-optic heaters for tuning. Moreover, power dissipation from such tuning can be avoided by permanently shifting the resonances via laser-based trimming.

\section{Results}

\begin{figure*}[htp!]
\centering
\includegraphics[width=1\textwidth]{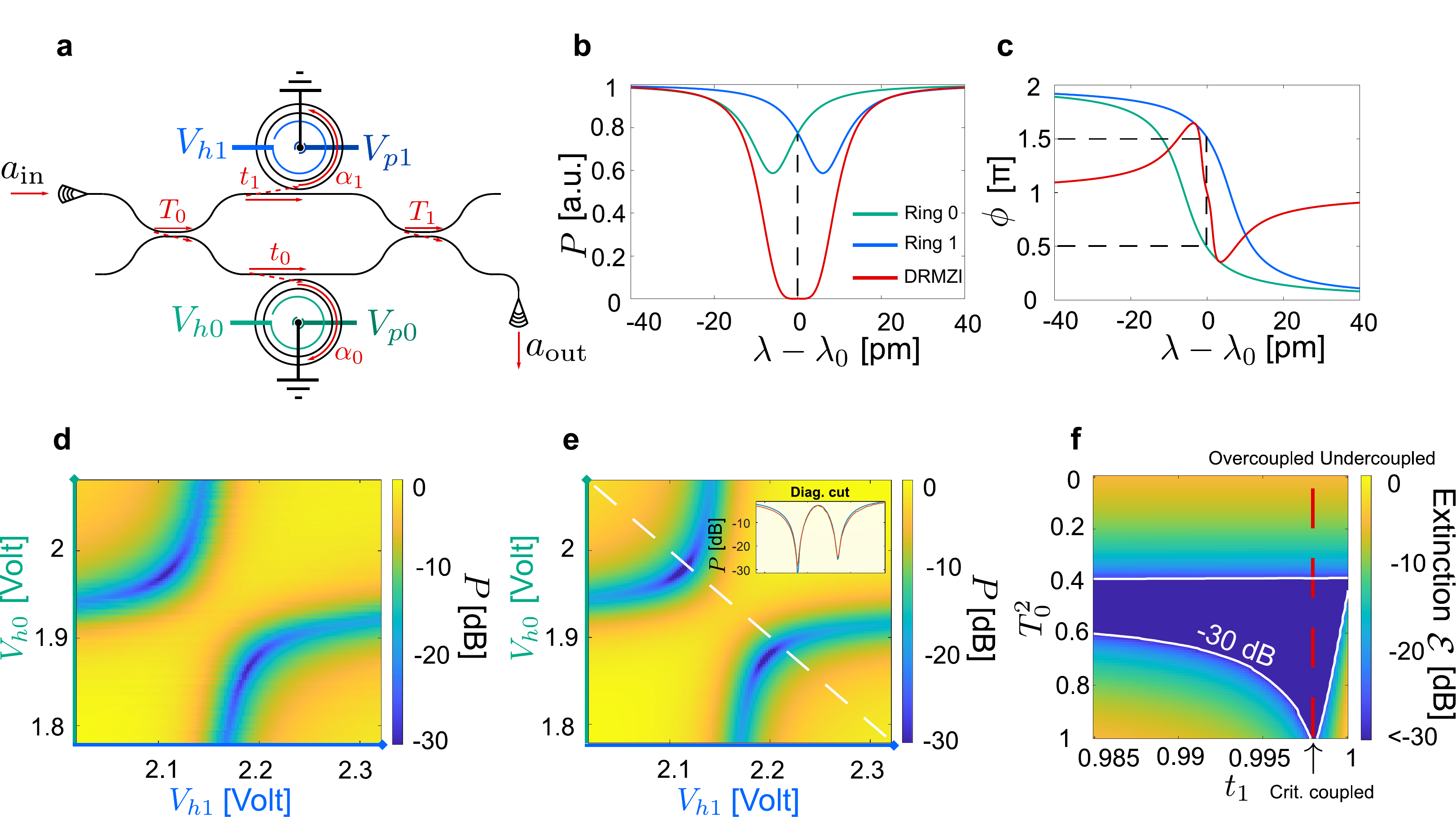}
\caption{Dual-ring-assisted Mach-Zehnder modulator (DRMZM). \textbf{a}, Schematic of a DRMZM. Input light amplitude is indicated as $a_\text{in}$, along with the transmissive amplitude coupling coefficients for beamsplitters ($T_{0,1}$) and ring-waveguide ($t_{0,1}$), and the piezo- $V_{p0,1}$ and heater- voltages $V_{h0,1}$ used for modulation and long-range tuning. $\alpha_{0,1}$ represents the single round trip ring transmission.
\textbf{b}, Transmitted powers $P=|a_\text{out}|^2$ around a target wavelength $\lambda_0$ for individual rings and the DRMZM. 
\textbf{c}, Phase $\phi=\text{arg}(a_\text{out})$ responses. These plots (\textbf{b},\textbf{c}) assume identical rings and beamsplitters. 
\textbf{d}, Measured normalised power $P=10\log_{10}({|a_{\text{out}}|^2/\text{max}(|a_{\text{out}}|^2))}$~dB as a function of heater voltages.
\textbf{d}, Fit of \textbf{e} to our model with the parameters: $\alpha_0=0.9962\pm1.7\cdot10^{-4}$, $\alpha_1=0.9975\pm1.9\cdot10^{-4}$, $t_0=0.983\pm1.8\cdot10^{-4}$, $t_1=0.9816\pm2\cdot10^{-4}$. $T_{0,1}^2=0.5$ are fixed. The $R^2$ value of the fit is 0.96. These values for $\alpha_{0,1}$ yield fitted intrinsic quality factors $Q_{i0}=2.7 \cdot 10^5\pm1.34\cdot10^4$ and $Q_{i1}=4 \cdot 10^5\pm3\cdot10^4$. The inset shows the profile along dotted diagonal line (red data, blue fit). \textbf{f}, Numerically extracted extinction achievable as a function of $T^2_0$ ($T_1^2=0.5$) and $t_1$ $(t_0=0.985)$. We set $\alpha_0=\alpha_1=0.998$. Blue region inside the contour indicates better than \SI{-30}{\decibel} extinction. }\label{fig:DRMZI}
\end{figure*}

Interfacing between the laser source and the APIC is a large programmable switch to uniformly distribute light into each DRMZM (Fig.~\ref{fig:archi}): in this case, implemented holographically via a commercial megapixel liquid crystal spatial light modulator (SLM)~\cite{christen2022integrated}. While such SLMs---with $\sim$100~\si{\hertz} update rate---cannot satisfy the speed criterion (C5) directly, they are ideal for static optical power fanout to balance light across the high speed DRMZMs in a scalable and reconfigurable way. By delegating the input light control to the millions of stable degrees-of-freedom on the SLM, we save the need for large-footprint, carefully-balanced, fabrication-sensitive splitter trees~\cite{Tao:08}.

Vertically radiating gratings couple the holographically-distributed light into and out of each DRMZM. We route the inputs and outputs of the modulators together in one corner of the APIC (Fig.~\ref{fig:archi}c), where we use a 1$\times$16 array of grating couplers for coupling in, and a two-dimensional 4$\times$4 array of grating couplers for coupling the modulated light out towards the target atomic systems.  Notably, the input and output grating couplers are oriented at \SI{90}{\degree} with respect to each other, such that a polarising beamsplitter (PBS) can be used to separate the in- and out- going light.
At the system level, a relevant characteristic is the optical power efficiency of each device $\eta = \text{max}_V(|a_{\text{out}}(V)|^2)/|a_{\text{in}}|^2$, where $a_{\text{in,out}}$ is the field amplitude at the input and output of a DRMZM and $V$ is applied voltage (Fig.~\ref{fig:DRMZI}a). 
Inefficiency is dominated by losses in the grating couplers. To boost efficiency, we take advantage of the aluminium layer underneath the grating which serves as the top piezo-electrode shown in Fig.~\ref{fig:archi}d. The reflective aluminium directs light upward and greatly enhances coupling efficiency~\cite{VanLaere:07}. We achieve an approximate per-grating efficiency of $\SI{30}{\percent}$ and a total throughput of $\eta\approx\SI{10}{\percent}$. 

\newpage % This is to make the pages come out nicely for arXiv submission.
The large channel count criterion (C2) motivates high optical channel density $\rho$. The resonant phase modulators used in the DRMZM afford a higher degree of compactness compared with conventional phase modulators. With an inter-device spacing of \SI{420}{\micro\metre} in our demonstrator cf. Fig.~\ref{fig:archi}b, we achieve a device density of approximately six devices per square-millimetre $\rho\sim6/\text{mm}^2$. If we use the entire reticle (\SI{2.2}{\centi\metre}$\times$\SI{2.2}{\centi\metre}) for our array, $\sim$2900 devices can be realised, with further improvements to compactness possible.

{\bf Performance metrics for quantum gates.} We now establish key performance metrics of our device. These metrics are guided by the application requirements for optical quantum control. The Rabi frequency of an optically driven atomic transition depends on the electrical field $\vec{E}(x,y,t)$ (transverse coordinates $x,y$) at the position of the atom. Experimentally, we measure the optical power using a photodiode $P(t)=\int I(x,y,t)dA$, where $I(x,y,t)\propto |\vec{E}(x,y,t)|^2$. 
For simplicity we consider a rectangular $\pi$ control pulse with constant light intensity $I_1$ in the ``on'' state.  The extinction is $\mathcal{E}=\frac{I_0}{I_1}$, where $I_0$ is the residual light intensity in the ``off'' state. We define the intensity normalised pulse error by $\Delta\mathcal{I}=\frac{\Delta I_1}{I_1}$. Given an intensity error $\Delta I_1$, the ``on'' state error $1-\mathcal{F}_1\sim(\Delta\mathcal{I})^2$ is proportional to the square of the intensity error. The ``off'' state error  $1-\mathcal{F}_0\sim\mathcal{E}$ scales linearly with extinction (see supplementary section I and \cite{nielsen2002quantum}). The requirements on $\mathcal{E}$ and $\Delta\mathcal{I}$ for low pulse error correspond to criteria (C3) and (C4). 

Criterion (C5) for fast switching times $\delta\tau$ follows from the need for many gate operations to be executed during the lifetime of the quantum state~\cite{bernien2017probing,Bruzewicz:19,Morgado:21}. Typical gate durations are in the range of tens to hundreds of nanoseconds for single-qubit rotations or entangling gates for cold atom systems~\cite{Levine:19,Morgado:21}, atom-like emitters in solids~\cite{Debroux:21}, and in the range of microseconds to milliseconds for trapped ion motional gates~\cite{Pino:21,Bruzewicz:19}.

{\bf Modulator architecture.} The DRMZM, schematically shown in Fig.~\ref{fig:DRMZI}a, lies at the core of our architecture. It consists of a Mach-Zehnder interferometer (MZI), with two 50:50 beamsplitters, and a ring resonator coupled to each arm of the interferometer. 

\begin{figure*}[htp!]
\centering
\includegraphics[width=1\textwidth]{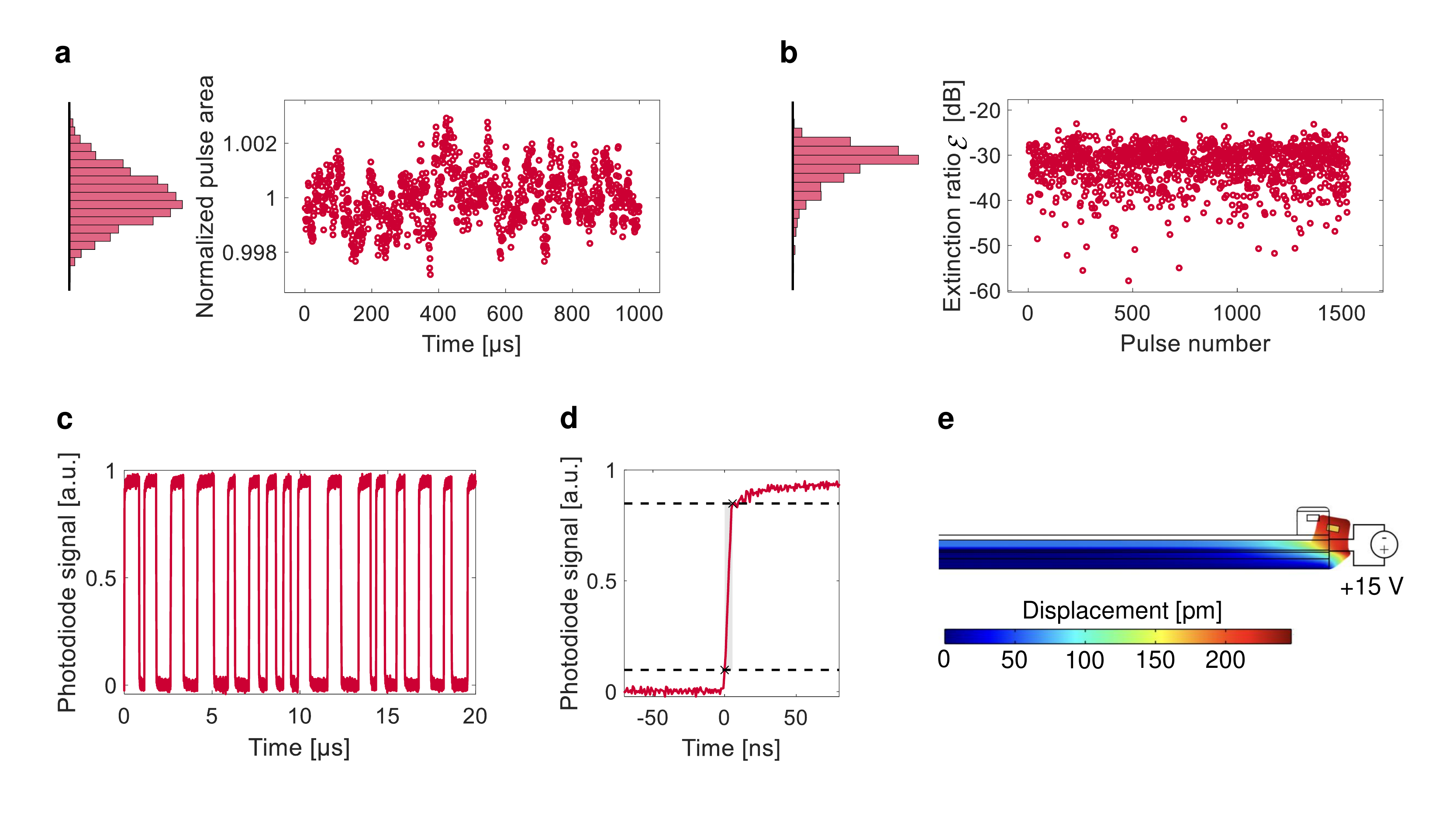}
\caption{Fast switching using piezoelectric actuators. \textbf{a}, Normalised pulse area vs time. Each point corresponds to a single pulse. Pulse area consistency ($1\sigma$) $\Delta\mathcal{I}=9.8\cdot 10^{-4}$ for a pulse train of \SI{1}{\milli\second} duration. This data uses duty cycle \SI{50}{\percent} and frequency \SI{1}{\mega\hertz}. \textbf{b}, Extinction ratio $\mathcal{E}$ for a random pulse train with pulse durations in between \SI{300}{\nano\second} and \SI{1000}{\nano\second}. \textbf{c}, Randomized pulse sequence. The optical output power in the ``on'' state is \SI{0.2}{\milli\watt}. \textbf{d}, Determination of the 10-\SI{90}{\percent} rise time $\delta\tau$. Black dashed lines indicate the \SI{90}{\percent} and \SI{10}{\percent} levels. \textbf{e}, Simulated ring waveguide displacement resulting from applied voltage of \SI{15}{\volt} to the bottom electrode of the piezo-stack. }\label{fig:pulse_trains}
\end{figure*}

Each ring resonator acts as a coupled phase and amplitude modulator with a well-known response function~\cite{bogaerts2012silicon}:
\begin{equation}
    a_{\text{SRout}}=a_{\text{SRin}}e^{i(\pi+\varphi)}\frac{\alpha-te^{-i\varphi}}{1-t\alpha e^{i\varphi}},
\end{equation}
where $a_{\text{SRin,out}}$ is the field amplitude at the input and output of a single ring, $\alpha$ is the attenuation coefficient for a single round trip in the ring, $t$ is the bus waveguide's self-coupling coefficient and $\varphi$ is the round trip phase.
For fast modulation, we rely on piezoelectric actuation cf. Fig.~\ref{fig:pulse_trains}e. In the overcoupled regime ($t<\alpha$), the ring acts as a ``force-multiplier'' for phase, where the small phase shifts possible through the fast but weak piezoelectric actuators can be amplified to a $\sim$$2\pi$ phase shift per ring (Fig.~\ref{fig:DRMZI}c)~\cite{gill2010ramzi}. Each ring is additionally equipped with a local heater (Fig.~\ref{fig:archi}d) for long range tuning to compensate for fabrication variations (as discussed in the next section).
Each overcoupled ring also modulates the amplitude in each arm of the MZI. By choosing the operating point of both rings correctly, we can select amplitude and phase in each arm to achieve in principle perfect destructive interference at the output port of the second beamsplitter. 
Furthermore, the two available degrees of freedom (phase tuning in both rings) afford full amplitude and phase control over the output electrical field (see supplementary section II).
This full field control is especially desirable in protocols where the phase of the optical field needs to be changed quickly~\cite{zhang2012fidelity}.

The simulated phase and amplitude response for both the isolated rings and the DRMZM is shown in Fig.~\ref{fig:DRMZI}b and c as a function of wavelength, assuming the ideal situation where the beamsplitters and rings are identical.
We achieve perfect extinction at the output port when the differential phase is $\pi$ and the light amplitude in both MZI arms are exactly balanced to match the amplitude splitting ratio of the out-coupling beamsplitter.
Fig.~\ref{fig:DRMZI}d shows the experimentally-measured output light power as a function of the heater voltages for each ring, which are used for long-range tuning to this ideal operating point.
The points of minimum power are clearly visible in both branches of the ``avoided crossing'' of the two resonances ($\mathcal{E}= \SI{-31}{\decibel}$ in top branch and \SI{-30}{\decibel} in bottom branch).
We fit the experimental data to a model with parameters defined in Fig.~\ref{fig:DRMZI}a. The result is shown in Fig.~\ref{fig:DRMZI}e, in good agreement with the measured data ($R^2=0.96$). 

DRMZM extinction---a critical figure-of-merit for (C3)---is robust to large fabrication variations, in contrast to regular MZMs. In MZMs, the principal limitation to the achievable light extinction is given by how well the two beamsplitters comprising the MZM are matched to each other. Unequal beamsplitter ratios in the DRMZM are manifest in the output power distribution (Fig.~\ref{fig:DRMZI}e) as a breaking of mirror symmetry along the diagonal.
The two points of minimum power translate along both branches as the difference between the beamsplitters increases. 
In Fig.~\ref{fig:DRMZI}f we show that a high extinction is achievable for a wide range of fabricated parameters.
In our case, we attribute the primary limitations to the measured extinction ratio to originate from small drifts of the resonance position during the measurement along with input polarisation misalignment.

\begin{figure}[htp!]
\centering
\includegraphics[width=0.47\textwidth]{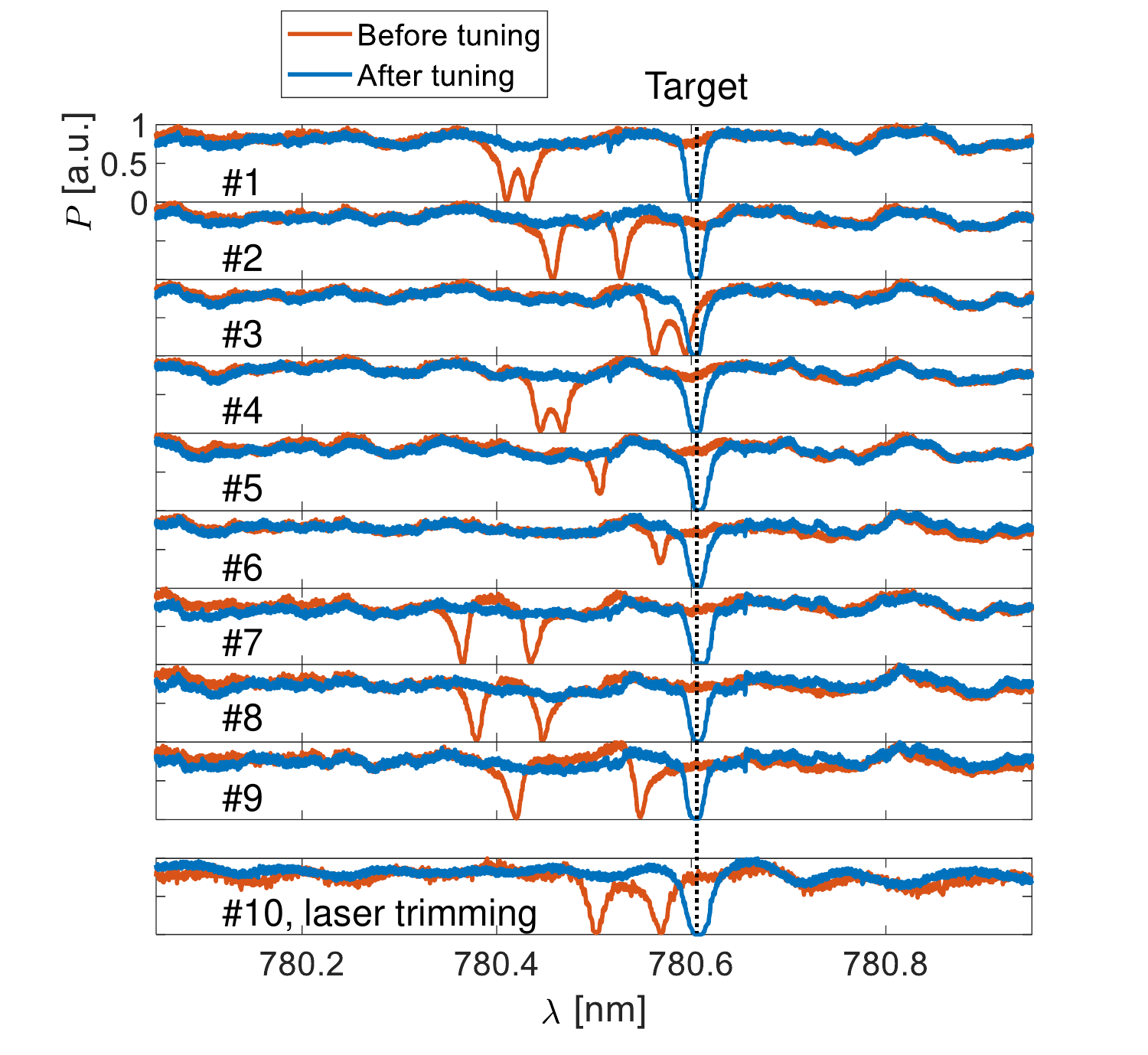}
\caption{Resonance tuning of DRMZMs. The top nine panels show the simultaneous alignment of the eighteen resonances of nine channels to a target wavelength (\SI{780.6}{\nano\metre}) using the integrated heaters. The bottom panel shows resonance alignment for a 10\textsuperscript{th} channel using nonvolatile laser trimming.
}\label{fig:spectrAlign}
\end{figure}

{\bf Resonance tuning.} 
Fabrication-induced variations in resonance frequency are a major issue in integrated photonic devices~\cite{Bogaerts:14}.
While our piezoelectric actuators are fast, their range is not sufficient to account for this fabrication-induced resonance spread.
In our system, as-fabricated resonances are distributed over hundreds of picometers or tens of linewidths, making modulation of a single laser frequency across the full array of devices impossible.
To retune the system to a common resonance, we developed a combination of tunable and nonvolatile resonance shifting based on (i) integrated heaters~\cite{Huang:20} and (ii) laser trimming~\cite{lipka2014hydrogenated,de2020laser,panuski2022full}. 
Fig.~\ref{fig:spectrAlign} plots the resonance positions for ten different DRMZMs before and after alignment.

For reconfigurable thermo-optic tuning, heating power is generated by a resistive coil in proximity of the waveguide, as shown in Fig.~\ref{fig:archi}d. 
We measure resonance tuning efficiencies of \SI{6}{pm/mW} (see supplementary section IV).
Negating fabrication spread requires only moderate powers, on the order of \SI{50}{\milli\watt} per ring.

To complement thermal tuning and relax requirements on dissipated power, we can permanently but non-reversibly shift resonances by trimming rings with a blue laser, notably without degradation of the optical quality factor (see supplementary section V). The accessible trimming range is larger than the observed variations in resonance positions, and operates with sufficient precision to target within the limited range of piezoelectric actuation.

{\bf Fast switching.} 
With resonances aligned at a desired operating point, we next demonstrate that our DRMZMs operate as fast light modulators with repeatable switching at high extinction ratios under piezoelectric actuation (C3-C5). Fig.~\ref{fig:pulse_trains}a plots the measured normalised individual pulse area for a \SI{1}{\mega\hertz} rectangular pulse train of \SI{1}{\milli\second} total duration. This measurement indicates a pulse area consistency with a ($1\sigma$) standard deviation of $\Delta \mathcal{I}=9.8\cdot 10^{-4}$ (C4). 
Fig.~\ref{fig:pulse_trains}d shows the fast switching behaviour of our device (C5). We observe a 10-\SI{90}{\percent} rise time of $(5.8\pm0.4)$~\si{\nano\second}.
Further, we demonstrate a high extinction ratio of $\mathcal{E} \sim \SI{-30}{\decibel}$ during switching (C3).
Fig.~\ref{fig:pulse_trains}b shows the extinction ratios for a pulse sequence with randomized pulse lengths (between \SI{300}{\nano\second} and \SI{1}{\micro\second}), as displayed in Fig.~\ref{fig:pulse_trains}c (see supplementary section III for eye diagrams).
Table \ref{tab} summarises key performance metrics.

\section{Discussion}
We introduced a SiN Atom-control Photonic Integrated Circuit (APIC) technology for scalable quantum control of atomic systems. Fabricated in a 200~mm wafer, CMOS-compatible process, our APICs achieve pulse errors of $9.8\cdot 10^{-4}$, extinction ratios $<$\SI{-30}{\decibel}, and rise times of $(5.8\pm0.4)$~\si{\nano\second}, enabling high precision optical manipulation of quantum states. Furthermore, we showed that the fabrication-induced variations in resonance frequencies can be compensated for by integrated heaters or nonvolatile laser trimming, an essential feature for large-scale resonant PICs. The CMOS-compatiblity of our fabrication process also enables us to manufacture our photonics architecture directly on top of a high-voltage CMOS driver~\cite{johansson2013review}.
There are no fundamental limits to scaling our approach to thousands of channels in such an architecture.  
While we demonstrated operation at around \SI{780}{\nano\metre}, our architecture is compatible with operation across the entire transparency window of SiN down to blue wavelengths.
Replacing SiN with wide-bandgap waveguiding materials such as aluminium oxide~\cite{West:19} or nitride~\cite{Wan:20} could enable operation down to UV wavelengths.

\newpage % This is to make the pages come out nicely for arXiv submission.
We anticipate that programmable optical control as achieved by our APIC platform will find widespread application for manipulating quantum systems, especially for atomic quantum control on arrays of neutral atoms~\cite{Graham:22,Ebadi:21,Pezzagna:21,Abobeih:22} or ions~\cite{Bruzewicz:19,Pino:21}. In simulated condensed matter systems~\cite{abanin2019colloquium} we anticipate that our architecture will enable experimental studies beyond global many-body effects and towards local phenomena such as topological defects and their associated quasi-particles~\cite{yan2021topological}.

\begin{center}
\begin{table}[h]
\begin{tabular}{ |c||c| c|c| }
\hline
Metric& Symbol& Value & Criterion  \\
\hline\hline
Wavelength &$\lambda_0$& \SI{780}{\nano\metre} & C1\\
\hline
Channel density &$\rho$& $6/\text{mm}^2$ & C2\\
\hline
Power efficiency &$\eta$ & \SI{10}{\percent}& C2 \\
\hline
Extinction &$\mathcal{E}$& $<\SI{-30}{\decibel}$& C3 \\ 
\hline
Infidelity ``off'' state &$1-\mathcal{F}_0$& $1.5\cdot10^{-3}$& C3   \\
\hline
Pulse error &$\Delta\mathcal{I}$& $1\cdot10^{-3}$& C4  \\  
\hline
Infidelity ``on'' state &$1-\mathcal{F}_1$& $2.5\cdot10^{-6}$& C4  \\ 
\hline
10-\SI{90}{\percent} rise time &$\delta\tau$& $(5.8\pm0.4)$~\si{\nano\second}& C5 \\
\hline
\end{tabular}
\caption{Performance metrics.}\label{tab}
\end{table}
\end{center}
\vspace{-1.5cm} % Hack to get the end of declarations to fit on this page.

\section{Methods}
\textbf{Device fabrication.} Our devices are fabricated at Sandia National Laboratories on \SI{200}{\milli\metre} silicon wafers using deep ultraviolet optical lithography~\cite{Stanfield:19}. The devices have three metal layers interconnected through vias: a routing layer (Al/Ti), a bottom electrode layer (Al) and a top electrode layer (Al). 
A piezoelectric AlN layer sandwiched inbetween the bottom and top electrode layers enables piezoelectric actuators. The bottom and top electrode layers are connected to bond pads through vias and the bottom metal routing layer. 
On top of the piezoelectric actuators, a SiN waveguide layer (\SI{300}{\nano\metre} thickness) is fabricated with silicon dioxide cladding above and below.
An amorphous silicon release layer fabricated below the piezoelectric actuators can be used to undercut devices via \ce{XeF2} etching~\cite{Stanfield:19}.
The devices presented in this work are not released, to allow for a larger modulation bandwidth due to the ``stiffer'' resulting actuator.
In Fig.~\ref{fig:DeviceProcess} the on-chip electrical routing is illustrated. 

\textbf{Packaging and drivers.}
After fabrication we dice the wafers into chips. The chips are glued on a copper block using a thermal epoxy. Wire bonds connect the chip to a printed circuit board (PCB). Fig.~\ref{fig:wirebond} shows a picture of the wire bonded chip.
The PCB connects our chip to the heater and piezoelectric actuator drivers.
The heaters are driven by Qontrol Q8iv modules delivering a maximum voltage of \SI{12}{\volt} and a maximum current of \SI{24}{\milli\ampere} (per channel). 
The piezoelectric actuators are driven by Spectrum M2p.6566-x4 arbitrary waveform generators with a maximum output rate of \SI{125}{MS/s} and an output level of $\pm\SI{3}{\volt}$ into \SI{50}{\ohm}.
This signal is amplified by a factor of five to $\pm\SI{15}{\volt}$ using high-bandwidth electronic amplifiers (Texas Instruments THS3491) on the PCB, fed by a $\pm\SI{16}{\volt}$ power supply. The maximum slew rate of the amplifiers is \SI{8000}{\volt\per\micro\second}.

\textbf{Actuation efficiency.}
To enhance resonance stability, we strongly overcouple the rings (effectively broadening the linewidth), while ensuring that each ring can be actuated to $\sim$$\pi/2$ phase over the voltage range of $\SI{30}{\volt}$  achieved by our amplifiers. 
Our measured actuation efficiency conforms with the previously demonstrated $\SI{0.4}{\pico\metre\per\volt}$~\cite{Stanfield:19}.
The driving voltage of our devices can be reduced by using enhanced piezo-materials such as scandium-doped AlN, which has a piezoelectric response that is up to 5 times larger than that of AlN~\cite{Teshigahara:12}. We can also choose to operate at lower driving voltages at the expense of a reduced light efficiency, as lower driving voltages result in smaller resonance shifts.

\section*{Author contributions}
AJM designed the photonic chips. AJM, DE and IC conceived the experiments. AJM, AH, TP and IC built the experimental setup. AJM and AH performed the data analysis and simulation. AH and AJM conducted the measurements with assistance from IC, TP, MD and HL. IC, TP and MZ developed the control software. MZ designed the electronic drivers. AJL fabricated the chips. HR performed numerical strain simulations. CL implemented the blue light trimming with assistance from AJM, IC and TP. AJM, AH, IC and DE wrote the manuscript with input from all authors. AJM, AH, GG, ME and DE supervised the project.\\
\textsuperscript{\textdagger}These authors contributed equally.

\begin{acknowledgments}
The authors acknowledge Christoper Panuski and Sivan Trajtenberg-Mills for contributions to the \verb+slmsuite+ software package \url{https://github.com/QPG-MIT/slmsuite}, along with Lilia Chan for designing the chassis and chip mount, and Kevin Dauphinais for designing the PCB of the electronic driver.
\end{acknowledgments}
\section*{Declarations}
AJM acknowledges support from the Feodor Lynen Research Fellowship of the Humboldt Foundation, the MITRE Moonshot program, the Defense Advanced Research Projects Agency ONISQ program and the Department of Energy Quantum Systems Accelerator (QSA). AH was supported by a Fellowship of the Belgian American Educational Foundation. 
IC acknowledges support from the National Defense Science and Engineering Graduate Fellowship Program and the National Science Foundation (NSF) award DMR-1747426. 
TP acknowledges support from the NSF Graduate Research Fellowship Program and the MIT Jacobs Presidential Fellowship.

\clearpage
\appendix

\bibliography{sn-bibliographyV2}% Produces the bibliography via BibTeX.

\begin{figure*}[htp!]
\centering
\includegraphics[width=0.8\textwidth]{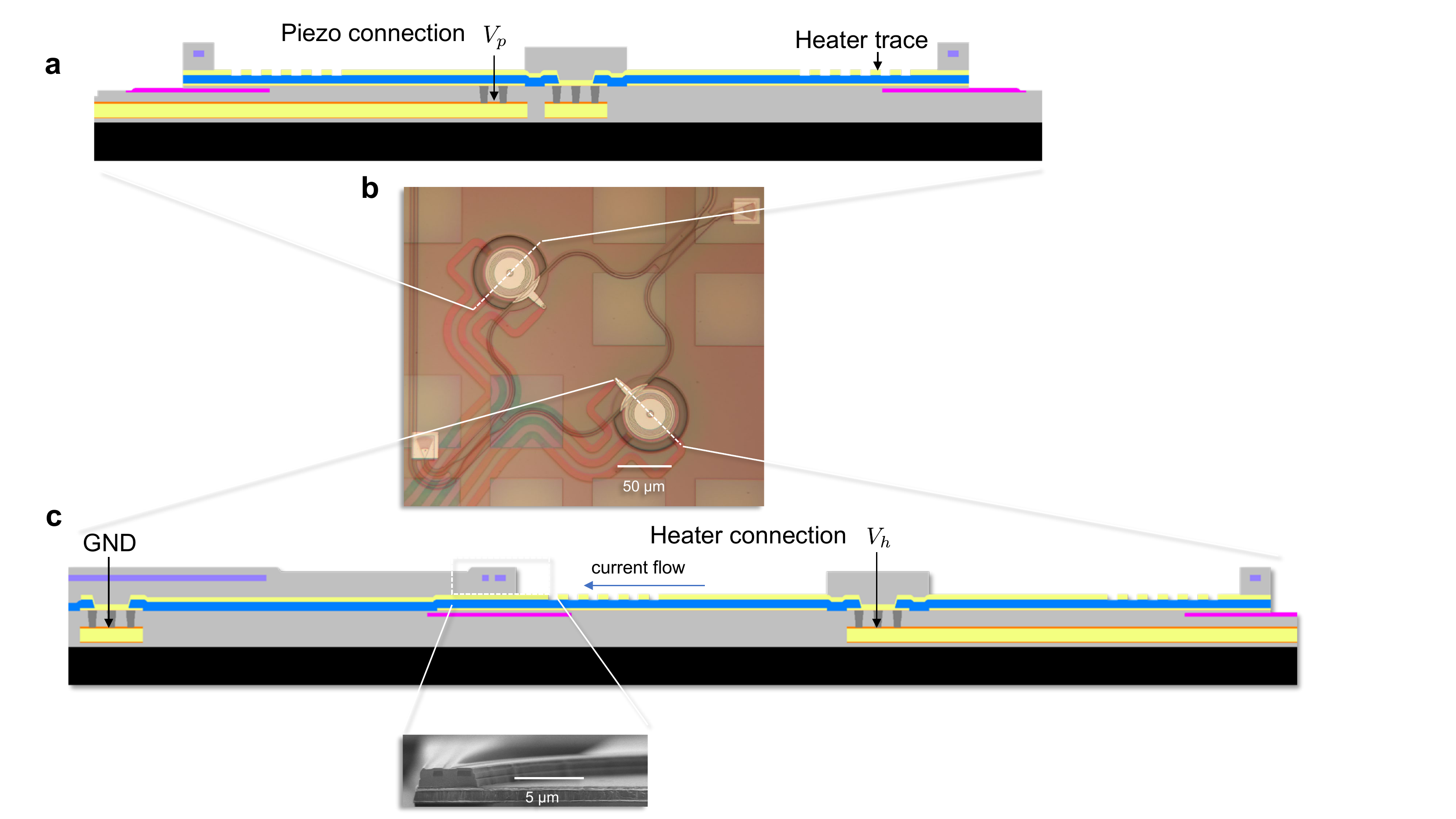}
\caption{We show two orthogonal cross sections through the wafer. \textbf{a}, Cut showing the connection to the bottom electrode of the piezo-actuator. Close to the waveguide we can see the cross-section of the heater spiral in the top metal. \textbf{b}, Microscope image illustrating the position for the cuts in \textbf{a} and \textbf{c}. \textbf{c}, Cut showing the ground (GND) and heater connections. The heating current flow from the heater terminal to ground via the high resistance heater spiral is shown. An SEM inset shows the region around the coupling waveguide.}\label{fig:DeviceProcess}
\end{figure*}

\begin{figure*}[h!]
\centering
\includegraphics[width=0.4\textwidth]{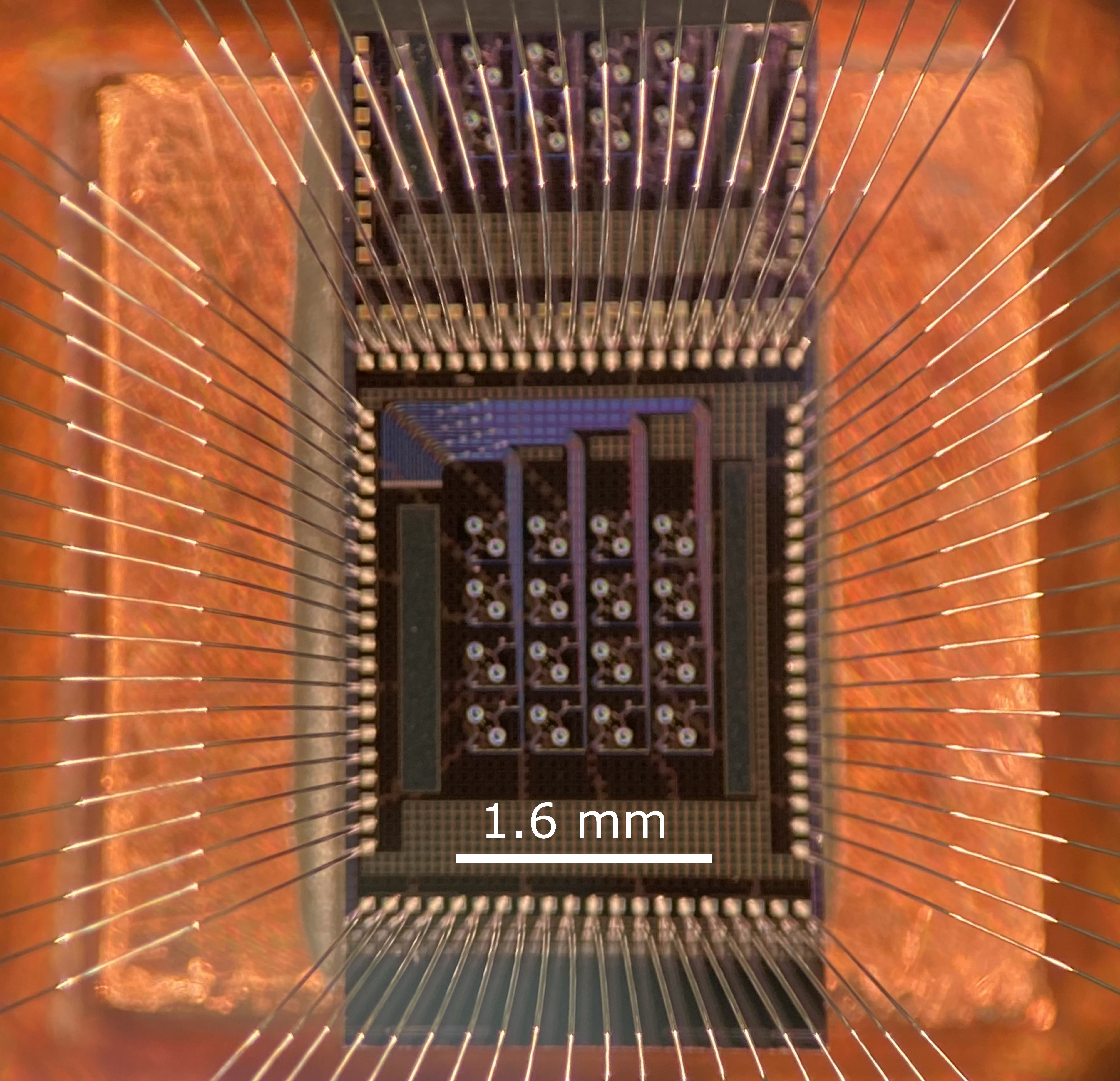}
\caption{Wire bonded chip on top of copper block.  We can see the individual devices arranged in a 4$\times$4 grid at the centre of the chip. Image credit: Merrimack Micro.}\label{fig:wirebond}
\end{figure*}

\end{document}

\preprint{APS/123-QED}

\title{Scalable photonic integrated circuits for programmable control of atomic systems: supplementary information}

\author{Adrian J Menssen\textsuperscript{\textdagger}}
\author{Artur Hermans\textsuperscript{\textdagger}}%
\affiliation{Massachusetts Institute of Technology, Cambridge, MA, USA}
\author{Ian Christen}
\affiliation{Massachusetts Institute of Technology, Cambridge, MA, USA}
\author{Thomas Propson}
\affiliation{Massachusetts Institute of Technology, Cambridge, MA, USA}
\author{Chao Li}
\affiliation{Massachusetts Institute of Technology, Cambridge, MA, USA}
\author{\\Andrew J Leenheer}%
\affiliation{Sandia National Laboratories, Albuquerque, NM, USA}
\author{Matthew Zimmermann}
\affiliation{The MITRE Corporation, Bedford, MA, USA}
\author{Mark Dong}
\affiliation{Massachusetts Institute of Technology, Cambridge, MA, USA}
\affiliation{The MITRE Corporation, Bedford, MA, USA}
\author{Hugo Larocque}
\affiliation{Massachusetts Institute of Technology, Cambridge, MA, USA}
\author{Hamza Raniwala}
\affiliation{Massachusetts Institute of Technology, Cambridge, MA, USA}
\author{Gerald Gilbert}
\affiliation{The MITRE Corporation, Princeton, NJ, USA}
\author{Matt Eichenfield}
\affiliation{Sandia National Laboratories, Albuquerque, NM, USA}
\affiliation{University of Arizona, Tucson, AZ, USA}
\author{Dirk R Englund}
\affiliation{Massachusetts Institute of Technology, Cambridge, MA, USA}

\maketitle

\date{\today}% It is always \today, today,

\section{Model of qubit gate fidelity}
We consider a simple model to estimate gate fidelity of optical gates on a general two-level system.
Considering pure states only, the fidelity of a quantum operation $\hat{U}(\epsilon)$ that is subject to an error $\epsilon$ is given by:
\begin{equation}
    \mathcal{F}(\hat{U}(\epsilon))=\text{min}_{\ket{\psi}}(|\braket{\psi|\hat{U}^\dagger(\epsilon) \hat{U}(0)|\psi}|^2).
\end{equation}
We consider the worst case scenario by minimising the quantum fidelity of the gate over all initial states $\ket{\psi}$.
For a single (atomic) two-level system, with infinite state lifetimes, subject to a resonant laser field, and assuming the rotating wave approximation, the evolution is given by the Hamiltonian:
\begin{equation}
   \hat{H}(\Delta\Omega)= \begin{pmatrix}
0 & (\Omega+\Delta\Omega)/2 \\
(\Omega+\Delta\Omega)/2 & 0 
\end{pmatrix},
\end{equation}
where $\Omega$ is the Rabi frequency and $\Delta\Omega$ is a (small) error in the Rabi frequency caused by intensity variations.
The corresponding unitary evolution is
$\hat{U}(\tau,\Delta\Omega)=e^{-i\hat{H}\tau}$.
We obtain for $\hat{U}(\tau,\Delta\Omega)$:
\begin{equation}
 \begin{pmatrix}
\phantom{-i}\cos[(\Omega+\Delta\Omega)/2\cdot \tau] & 
-i\sin[(\Omega+\Delta\Omega)/2\cdot \tau] \\
-i\sin[(\Omega+\Delta\Omega)/2\cdot \tau] & 
\phantom{-i}\cos[(\Omega+\Delta\Omega)/2\cdot \tau]
\end{pmatrix}.
\end{equation}

We can express the quantum gate fidelity in terms of an error in the phase of the unitary evolution $\Delta\phi=\Delta\Omega \tau/2$.
The minimum fidelity is to third order in $\Delta\phi$ given by:
\begin{equation}
    \mathcal{F}=1-(\Delta\phi)^2+\mathcal{O}((\Delta\phi)^4).
\end{equation}
The phase error for a $\pi$-pulse is given by:
\begin{equation}
    \Delta\phi=\frac{\Delta\Omega_1}{2}\tau_\pi=\frac{1}{2}\frac{\partial \Omega_1}{\partial I_1}\Delta I_1\cdot\frac{\pi}{\Omega_1}=\frac{\Delta I_1}{I_1}\frac{\pi}{4},
\end{equation}
with $\tau_{\pi}=\frac{\pi}{\Omega_1}$ and $\Omega_1=c\sqrt{I_1}$.
For the ``off'' state the phase-error is given by:
\begin{equation}
    \Delta\phi
    =\frac{\pi}{2}\sqrt{\frac{I_0}{I_1}}.
\end{equation}
We finally obtain for the ``on'' and ``off'' state fidelities:
\begin{eqnarray}
\mathcal{F}_{1}&=&1-\left(\frac{\pi}{4}\right)^2\left(\frac{\Delta I_1}{I_1}\right)^2\\
\mathcal{F}_{0}&=&1-\left(\frac{\pi}{2}\right)^2\frac{I_0}{I_1}.
\end{eqnarray}
Note that these fidelities will be different for other kinds of protocols, e.g. when considering a two photon transition.

\begin{figure}[htbp]
\centering
\includegraphics[width=0.5\textwidth]{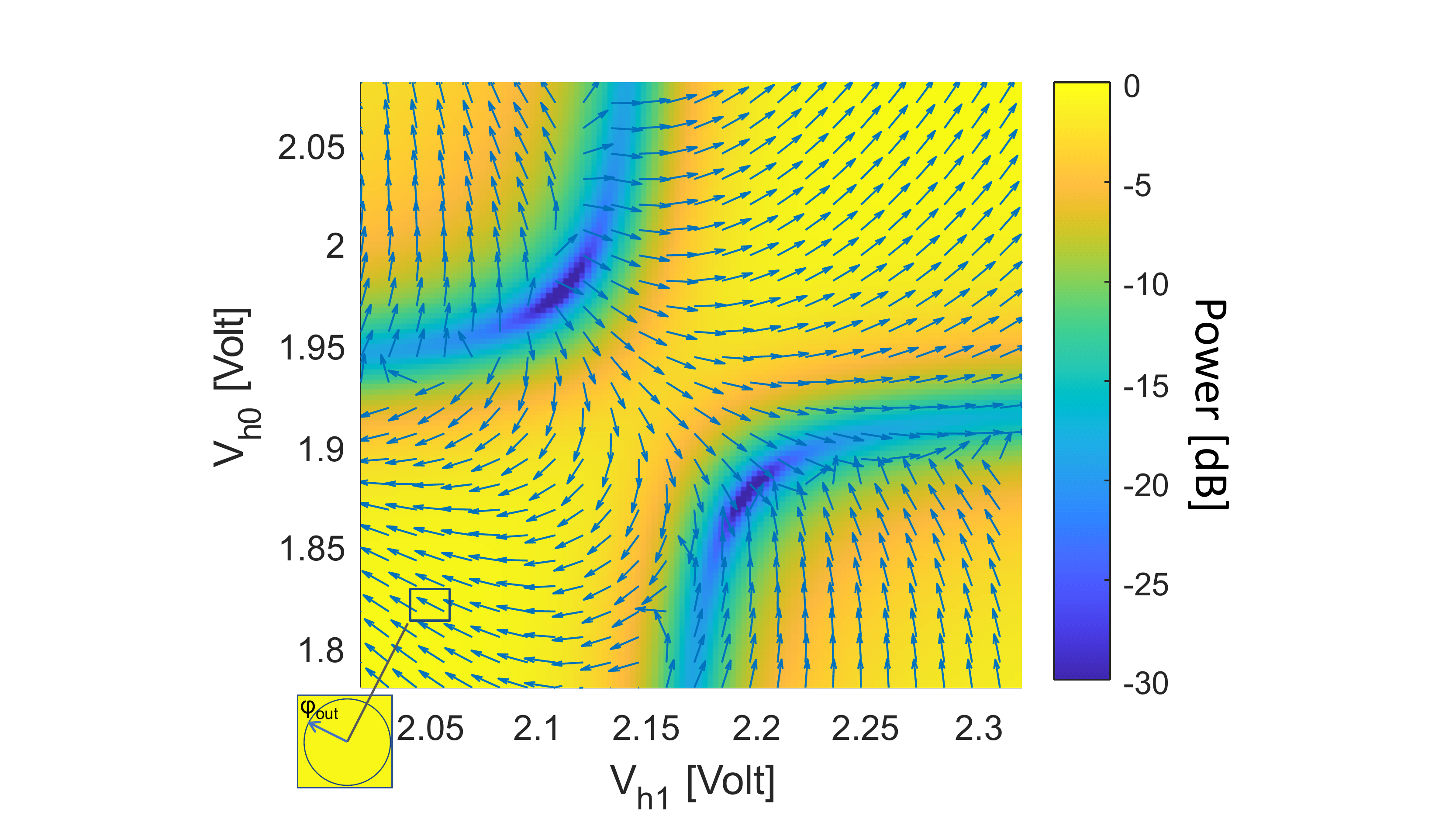}
\caption{Simulated phase and power response of dual-ring-assisted Mach-Zehnder modulator (DRMZM). Vector field illustrates phase of the output light at each point.}\label{fig:phase}
\end{figure}
\section{Phase and amplitude response}
The starting point of our DRMZI model is the response of a single ring
\begin{equation}
D(\alpha,t,\varphi)=e^{i(\pi+\varphi)}\frac{\alpha-te^{-i\varphi}}{1-t\alpha e^{i\varphi}},
\end{equation}
where $\varphi_i=2\pi L(n_0+dn_i)/\lambda$, with $L$ the ring circumference, $\lambda$ the wavelength, $n_0$ the refractive index and $dn_i$ a small index change introduced by the actuation.
The response of two rings in the basis of the two optical modes of the MZI is then given by:
\begin{equation}
\hat{D}=\begin{pmatrix}
D(\alpha_0,t_0,\varphi_0)e^{i\phi} & 0 \\
0 & D(\alpha_1,t_1,\varphi_1) 
\end{pmatrix}.    
\end{equation}
$\phi$ is a constant differential phase between the two arms of the MZI. 
We then construct the response of the DRMZI by combining the unitary matrices representing the two beamsplitters $\hat{U}_{\text{BS}}(T)$ with the response functions of the two rings
\begin{equation}
\vec{a}_{\text{out}}=\hat{U}_{\text{BS}}(T_1)\cdot\hat{D}\cdot\hat{U}_{\text{BS}}(T_0)\cdot\vec{a}_{\text{in}}.
\end{equation}
We then assume that we send light into the first input $\vec{a}_{\text{in}}$ and measure at the second output $\vec{a}_{\text{out}}$ (the cross port), where $a_{\text{rej.}}$ is the rejected light amplitude:
\begin{equation}
\vec{a}_{\text{in}}=\begin{pmatrix}
a_{\text{in}} \\
0 
\end{pmatrix},
\,\,\,\,\,\,\,\,\,\,
\vec{a}_{\text{out}}=\begin{pmatrix}
a_{\text{rej.}} \\
a_{\text{out}} 
\end{pmatrix}.
\end{equation}
In Fig. \ref{fig:phase} we show the phase and power response of the device as a function of the actuation voltages. The vector field plot indicates the value of the phase at each point. We note that the two minima are singularities and the phase-vector field forms vortices around both.

\begin{figure}[htbp]
\centering
\includegraphics[width=0.42\textwidth]{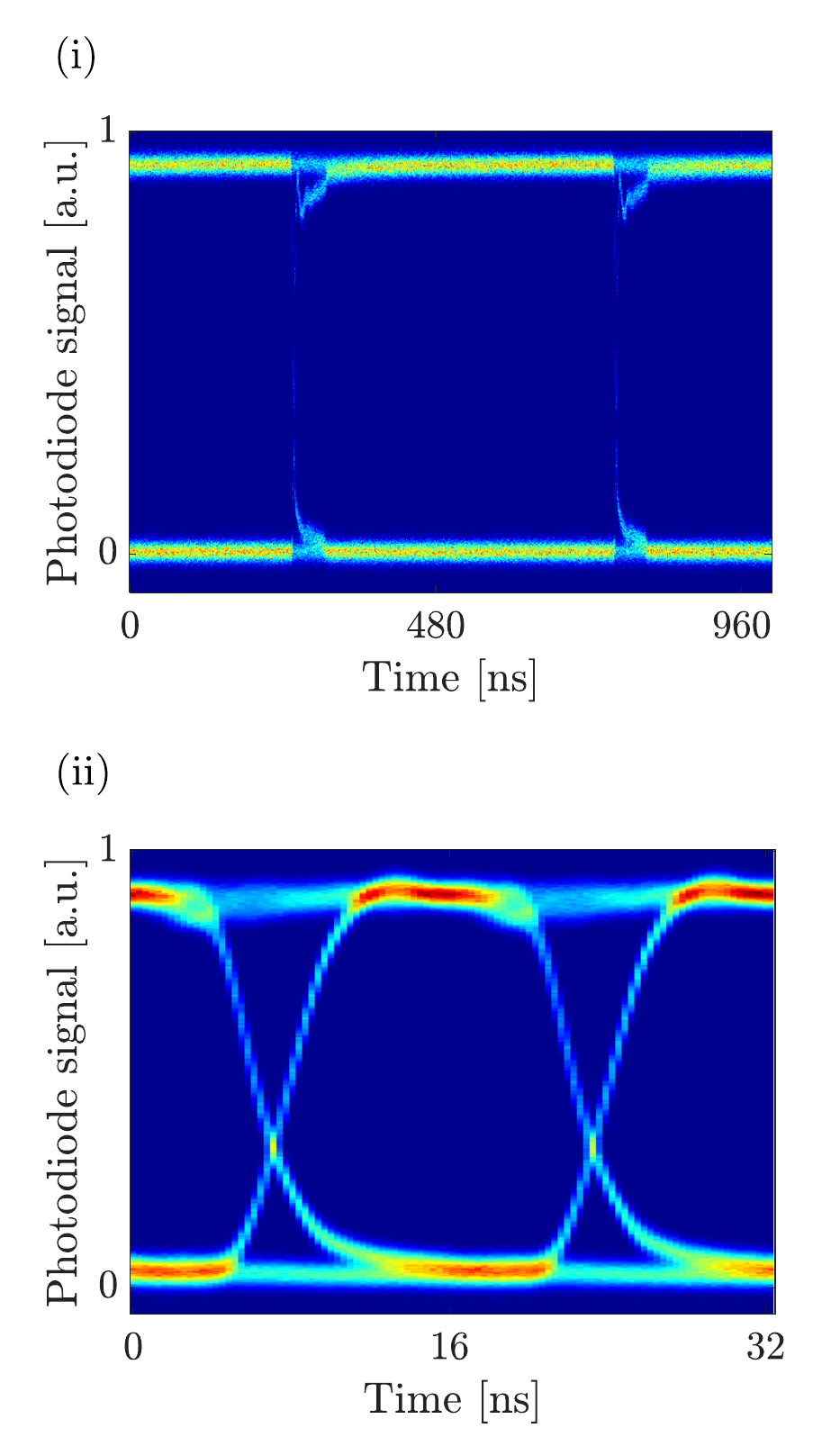}
\caption{Eye diagrams for a symbol duration time of (i) \SI{504}{\nano\second} (sample interval of oscilloscope set to \SI{1.6}{\nano\second}) and (ii) \SI{16}{\nano\second} (sample interval of oscilloscope set to \SI{0.32}{\nano\second}). The total duration of the randomized pulse trains is \SI{1}{\milli\second}.}\label{fig:eye}
\end{figure}

\section{Fast switching}
Fig. \ref{fig:eye} shows the eye diagrams for two different symbol duration times. For these measurements, we first set the heater voltages of our DRMZM to obtain minimum transmission. Next we apply a \SI{1}{\milli\second} long randomized pulse train to the piezoelectric actuators to switch between the “off” and “on” states. The eye diagrams are constructed by triggering on the rising edges of a clock signal at the symbol rate.

\section{Thermo-optic tuning and heater characteristics}
\begin{figure}[htp!]
\centering
\includegraphics[width=0.43\textwidth]{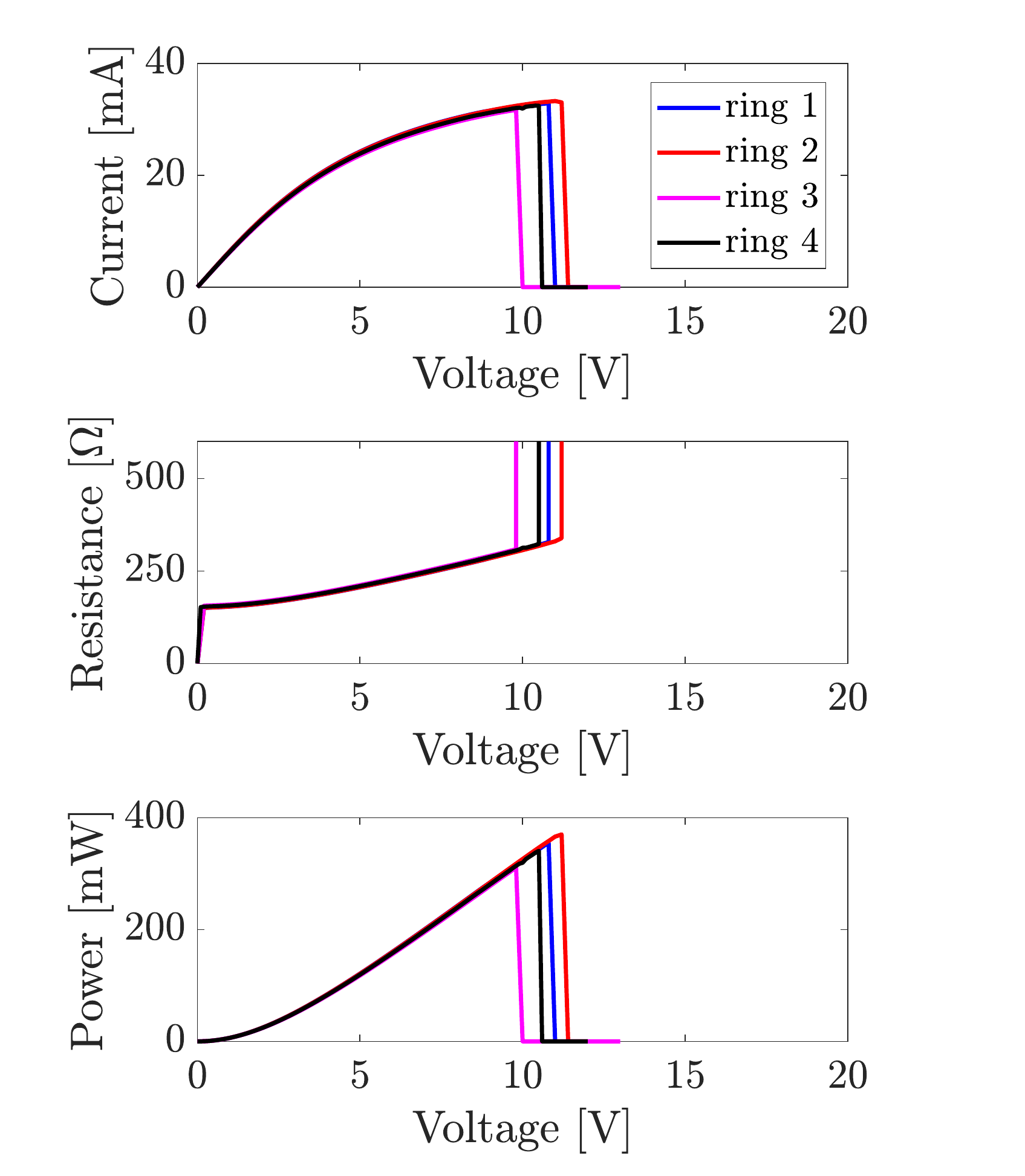}
\caption{Thermal actuation of several rings (radius of \SI{20}{\micro\metre}). The increase of resistance is consistent with the behavior of aluminium at high temperatures.}\label{fig:destr}
\end{figure}

\begin{figure}[htp!]
\centering
\includegraphics[width=0.5\textwidth]{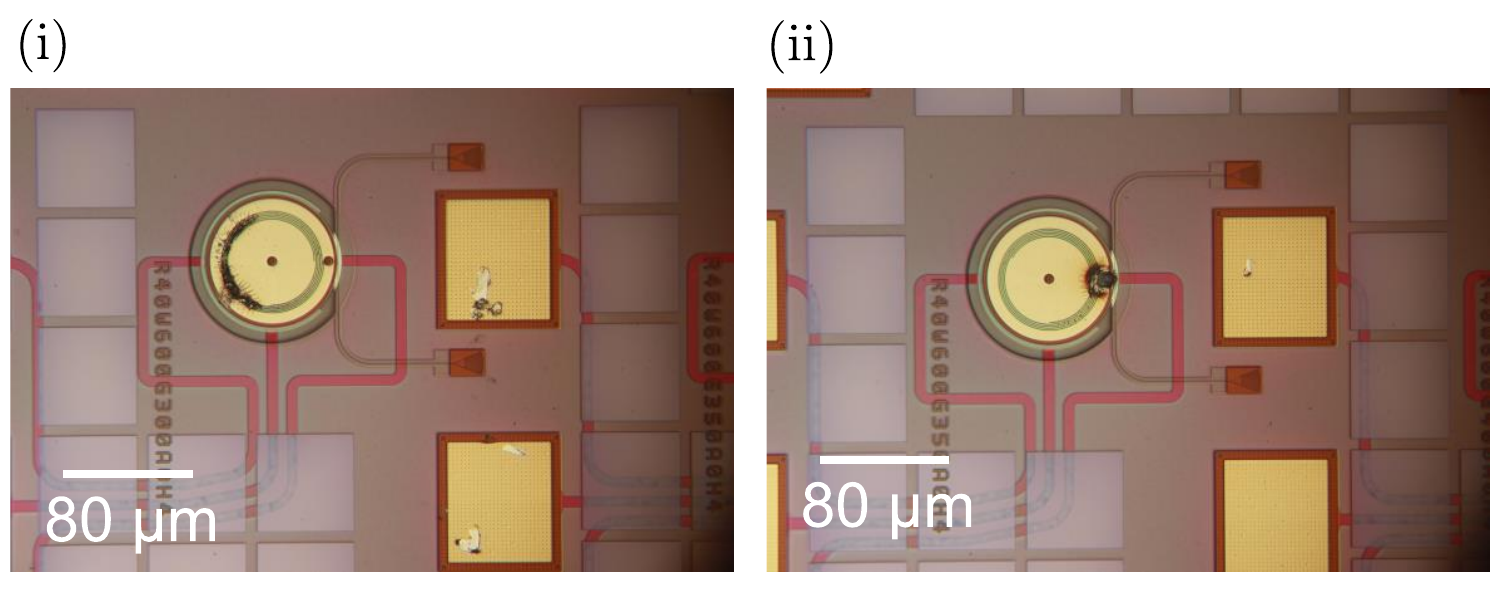}
\caption{Two failure modes of a single test ring. (i) Damage to the heater trace itself. (ii) Damage to the via.}\label{fig:damage}
\end{figure}

In Fig. \ref{fig:destr} we show the response curves for four separate ring heaters, plotting current, resistance, and electrical power versus applied control voltage. All rings have similar behaviour up to their breaking point at around \SI{300}{\milli\watt} (where the current suddenly drops). Nevertheless, this \SI{300}{\milli\watt} range is much larger than the ranges which we generally tune with the heaters.
In Fig. \ref{fig:damage}, two failure modes for individual test rings are shown. In Fig. \ref{fig:damage}(i) the heater trace on the left side of the ring shows significant damage, while in Fig. \ref{fig:damage}(ii) a via connection suffered damage. 

\begin{figure}[htp!]
\centering
\includegraphics[width=0.5\textwidth]{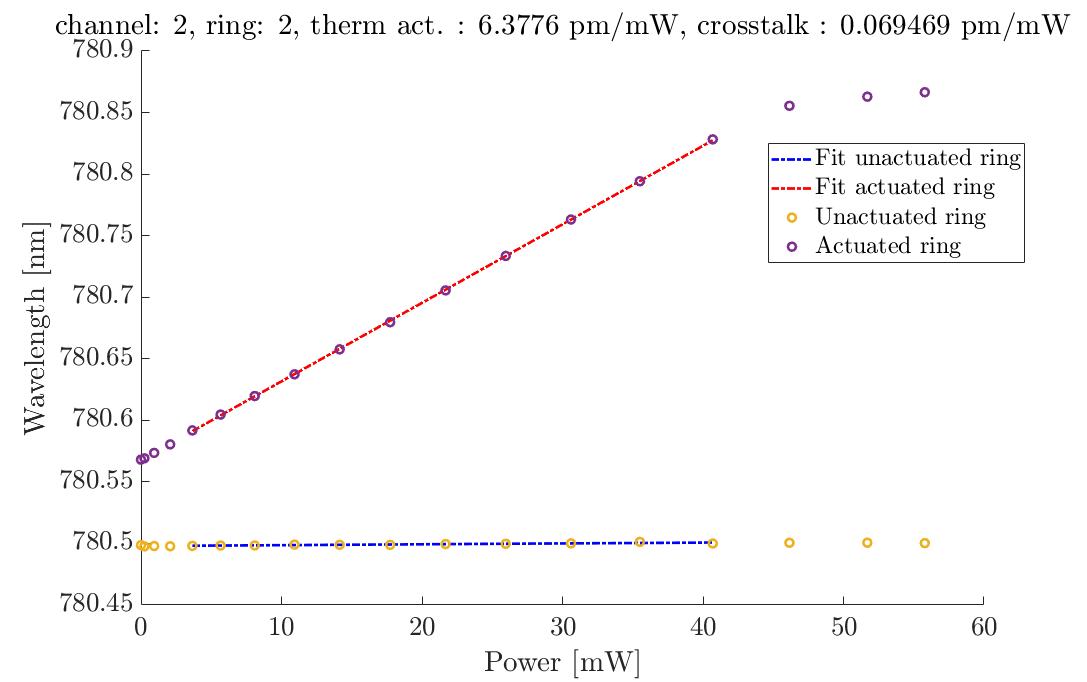}
\caption{Resonance position vs heating power for the rings in a DRMZM.}\label{fig:thermalAcct}
\end{figure}

In Fig. \ref{fig:thermalAcct} we show the wavelength shift as a function of heating power for the DRMZM rings. For each power value, the laser is swept in wavelength across the resonances to fit and record the resonance positions.
In the range over which we thermally actuate the rings, the wavelength shift is linear in power. The plateauing at powers greater than $\sim\SI{45}{\milli\watt}$ occurs as the single channel current limit of our current source is reached. We measure $\sim6.4$~pm/mW actuation and around $-20$~dB thermal crosstalk to the neighbouring ring.

\section{Laser trimming}
\begin{figure*}[htbp]
\centering
\includegraphics[width=0.95\textwidth]{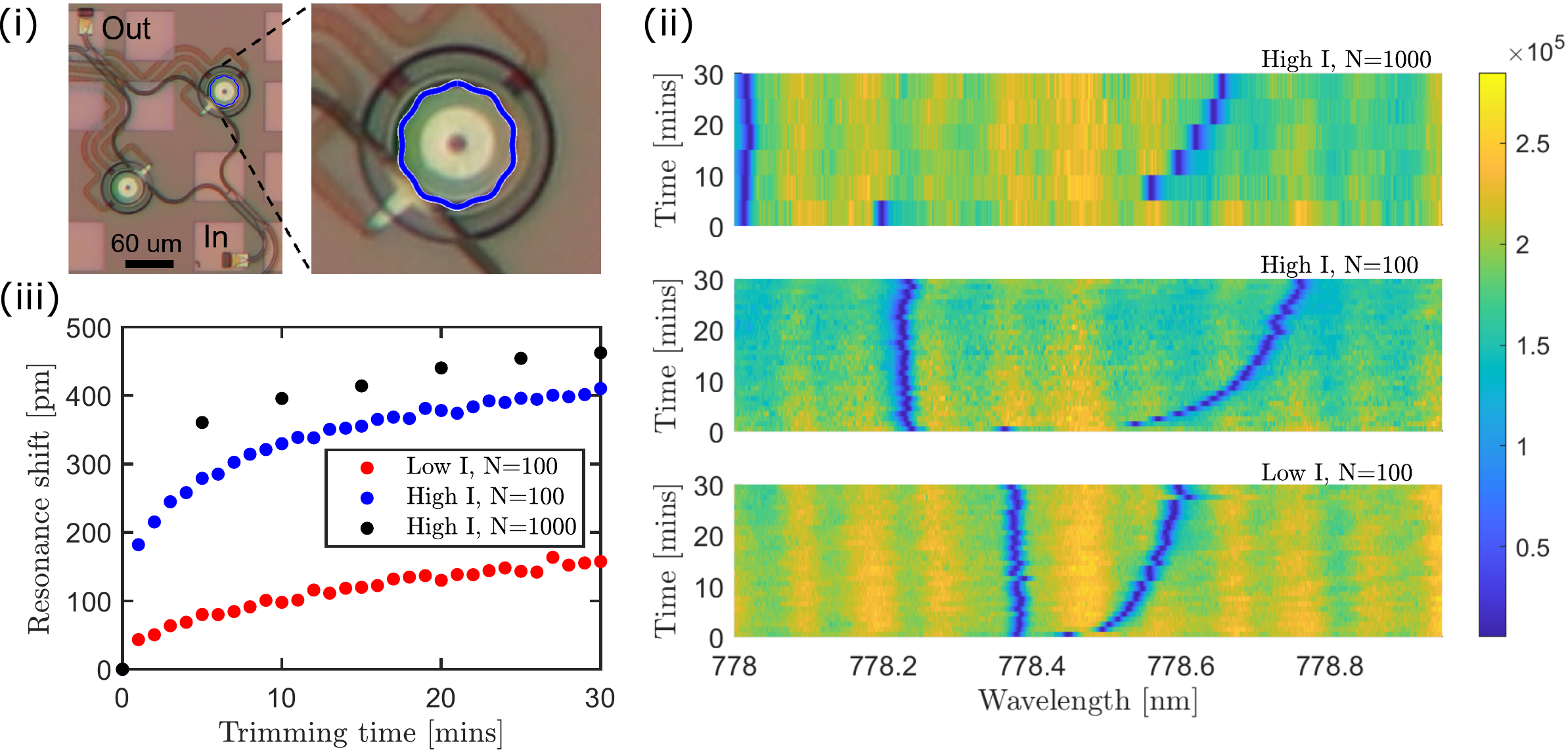}
\caption{Laser trimming characterization.
(i) A microscope image of the DRMZM. The focused trimming beam follows the blue circle with dithered radius overlapped with the silicon nitride ring.
(ii) A collection of measurements for the laser-trimmed resonance across multiple modulators on a separate off-PCB chip. The false color [a.u.] represents the output power. The trimming duration per data set is five minutes for the upper panel and one minute for the lower two panels.
(iii) The resonance shift versus trimming time, analysed from (ii). $N$ is the number of trimming steps per round trip. $I$ is the trimming laser intensity $I = 0.1~(0.5) ~\si{\milli\watt\per\micro\meter\squared}$ for low (high) intensity, respectively.
}\label{fig:shiftvstime}
\end{figure*}

To accomplish laser-based permanent shifting of ring resonances, we introduce an additional path to controllably inject a trimming laser onto each ring.
This path includes a blue CW laser (Cobolt 06-MLD 405 nm) and, critically, another SLM for controllable beam focusing and positioning (Thorlabs EXULUS-HD2HP). 
This path is merged into the existing fanout and characterization paths by using a long-pass dichroic filter directly before the objective (Thorlabs RMS10X). The setup delivers trimming power up to 40 mW transmitted through the objective with a measured efficiency of 74\% diffracted towards the first order mode, which is used for trimming. The focused beam radius at $1/e^2$ intensity is measured to be \SI{4}{\micro\metre} at the plane of the chip.
Tuning is accomplished by scanning this SLM-programmable first order beam across the waveguide in a dithered circular pattern as shown in Fig.~\ref{fig:shiftvstime}(i): a circle with a sinusoidal perturbation to radius to compensate for any misalignment between the center of the trimming path and the center of the targeted ring. 
We discretise this path into $N$ steps around the circle.
Every iteration step along the circle takes $\sim$ 0.25 seconds, and $N$ varies with experiment according to Fig.~\ref{fig:shiftvstime}(iii).
Once the laser intensity and the SLM scanning parameters are set, we leave the trimming light constantly on. We then periodically switch a shutter in the trimming path to expose the SiN ring for either one minute or five minutes. 

At the end of each exposure, we measure the two resonances of the DRMZM shown in Fig.~\ref{fig:shiftvstime}(ii). In these examples, the resonance to the left is unperturbed (with small variations stemming from temperature fluctuations) and serves as the reference, while the other is red-shifted in wavelength due to laser trimming. 
These results show that this post-fabrication trimming is sufficiently wide to compensate for standard observed resonance spread in our 4$\times$4 device arrays ($\sim$0.4 nm). 
To target the devices for a specific wavelength we select a chip with resonances close to the desired wavelength. This minimises the required trimming power/duration. 
In addition, we observe no significant change in the extracted full width at half maximum (inversely proportional to quality factor $Q$) given the data in Fig.~\ref{fig:shiftvstime}(ii).
We measure the laser-trimmed devices again after roughly one month of ambient storage and observe no relaxation in trimming. The heaters and piezo actuators underneath the ring resonators operate after the laser trimming.

Note that while the results here are obtained at 405~nm with a fiber-coupled source, the laser trimming results reported in the main text use a low-cost 445~nm laser diode (Nichia NDB7A75). The diffraction-limited mode size of the fiber-coupled source allowed us to calibrate these supplementary trimming results. We operate the 445 nm diode at 1~A and deliver 420~mW to the first-order, but achieve an averaged intensity of only $\sim$ 0.6 $\si{\milli\watt\per\micro\meter\squared}$ due to its larger spatial mode.

\begin{figure*}[htbp]
\centering
\includegraphics[width=1\textwidth]{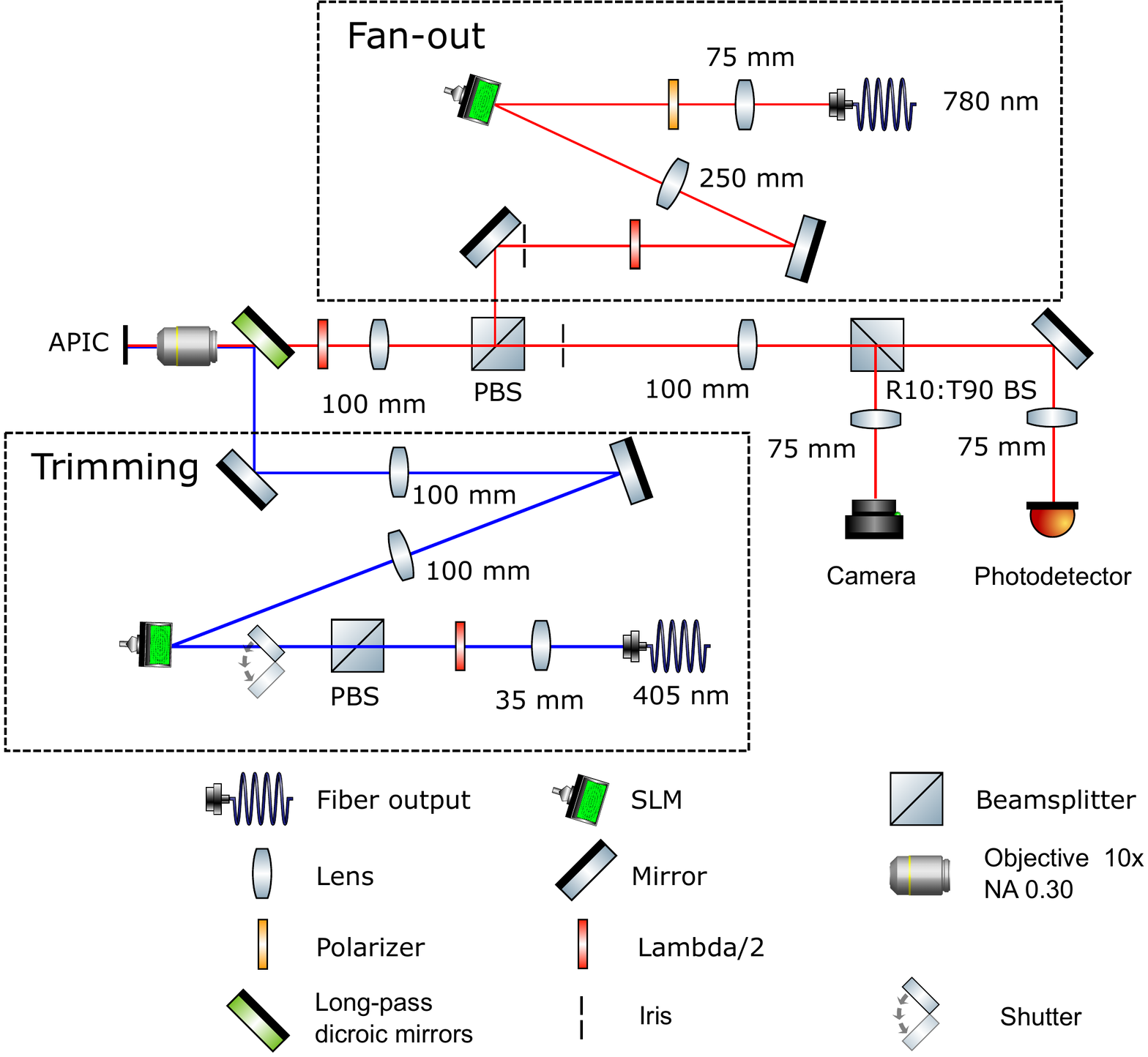}
\caption{Optical setup. The upper section, with an SLM in the Fourier domain, produces the fanout hologram to couple input beams to devices. These input beams are imaged onto the APIC chip by an objective. The polarisation of the modulated light is rotated by 90 degrees (by virtue of grating orientation) and transmitted through to the imaging portion, where a camera is used for slow feedback and a photodiode is used for characterising the fast signal response. 
In the lower section we show the 405~nm trimming setup. 
}\label{fig:setup}
\end{figure*}